\documentstyle[12pt]{article}

\def  \be	 {\begin{equation}}
\def  \ee	 {\end{equation}}
\def  \beq	 {\begin{eqnarray}}
\def  \eeq	 {\end{eqnarray}}

\def  \rlarrows	 {{\lower.6ex\hbox{$\;\buildrel{\mbox{\footnotesize$\longrightarrow$}}
		 \over{\mbox{\footnotesize$\longleftarrow$}}\;$}}}

\begin{document}

\title{\bf On the Internal Structure of Relativistic Jets\footnote{Astronomy Letters, 2000, 
{\bf 26}, (4), 208--218, Translated from Russian by V. Astakhov}}

\author{Vassily S. Beskin$^1$ and Leonid Malyshkin$^2$\\
{\it\small $^1$Lebedev Physical Institute, Russian Academy of Sciences, Leninskii pr. 53, Moscow 117924, Russia} \\
{\it\small $^2$Princeton University Observatory, Peyton Hall, Princeton, NJ 08544}\\
{\small $^1$beskin@lpi.ru, $^2$leonmal@astro.princeton.edu}}

\date{}

\maketitle

\begin{abstract}
A magnetohydrodynamic model is constructed for a
cylindrical jet immersed in an external uniform magnetic field. It is
shown that, as in the force-free case, the total electric current within
the jet can be zero. The particle energetics and the magnetic field
structure are determined in a self-consistent way; all jet parameters
depend on the physical conditions in the external medium. In
particular, we show that a region with subsonic flow can exist in the
central jet regions. In actual relativistic jets, most of the energy is
transferred by the electromagnetic field only when the magnetization parameter
is sufficiently large, $\sigma>10^6$. We also show that, in general,
the well-known solution with a central core, $B_z = B_0/(1+\varpi^2/\varpi_c^2)$, 
cannot be realized in the presence of an external medium.
\end{abstract}


\section{Introduction}

The formation mechanism of jets is a key issue in the study of the
magnetospheric structure of compact astrophysical objects. Indeed, jets
are observed in most compact sources, ranging from active galactic
nuclei (AGNs), quasars, and radio galaxies~\cite{1} to accreting neutron
stars, solar-mass black holes (SS 433, X-ray novae)~\cite{2}, and young
stellar objects~\cite{3}. Moreover, jets have also been recently discovered
in young radio pulsars~\cite{4, 5}. At the same time, in most studies devoted
to the magnetohydrodynamic (MHD) model of such objects~\cite{6,7,8,9,10,11}, in which
the formation of jets is coupled with the attraction of longitudinal
currents flowing in the magnetosphere, the attention was focused on
intrinsic collimation in the sense that the effect of the external
medium was assumed to be marginal. However, such a situation is possible
only for a nonzero total current $I$ flowing within the jet~\cite{12}, so the
question of its closure in the outer parts of the magnetosphere arises.
On the other hand, the longitudinal current is often constrained by the
regularity condition at the fast magnetosonic surface, which by no means
always leads to the sufficiently large longitudinal currents required
for collimation~\cite{13}. In other words, with the exception of the
force-free case~\cite{14}, as yet no working model of a jet in which, on the
one hand, the total electric current would be zero and, on the other
hand, the total magnetic flux $\Psi_0$ in the jet would be finite, has
been constructed. However, the force-free approximation, in which, by
definition, the particle energy density is disregarded, does not allow
the fraction of energy transferred by the outflowing plasma to be
determined.

At the same time, the question of collimation cannot be solved in
isolation from the external conditions (see, e.g.,~\cite{15, 16}). In
particular, this is clear even from a popular example of the
magnetosphere of a compact object with a monopole magnetic field,
because for any arbitrarily weak external regular magnetic field, the
monopole solution (for which the magnetic field falls off as $r^{-2}$)
cannot be extended to infinity. Moreover, as is well known from an
example of moving cosmic bodies, such as Jupiter's moons~\cite{17} or
artificial Earth satellites~\cite{18, 19}, as well as radio pulsars~\cite{20}, the
external magnetic field can serve as an effective transfer link, which
occasionally determines the general energy losses of the system. For
this reason, constructing a consistent magnetospheric model for compact
objects immersed in an external magnetic field is, in our view, of
undeniable interest, especially since, as was noted above, such a jet
model was previously constructed in the force-free approximation~\cite{14}.

Undoubtedly, the existence of an external regular magnetic field in the
vicinity of compact objects is largely open to question. The regular
magnetic field in our Galaxy, i.e., the field that is constant on scales
comparable to the sizes of our Galaxy, is known to be
\begin{equation}
B_{\rm ext} \sim 10^{-6}\,{\rm G},
\label{1}
\end{equation}
and essentially matches the random magnetic-field component, which
varies even on scales of several parsecs~\cite{21}. However, if the
collimation is assumed to be actually produced by an external magnetic
field, it becomes possible to estimate the jet radius. Indeed, assuming
the magnetic field in the jet to be similar to the external magnetic
field~(\ref{1}), we obtain from the condition for the conservation of magnetic
flux
\begin{equation}
r_{\rm j} \sim R\left(\frac{B_{\rm in}}{B_{\rm ext}}\right)^{1/2},
\label{2}
\end{equation}
where $R$ and  $B_{\rm in}$  are the radius and magnetic field of the compact
object, respectively. For example, for AGNs ($B_{\rm in}\sim 10^4\,{\rm G}$, 
$R\sim 10^{13}\,{\rm cm}$), we have
\begin{equation}
r_{\rm j} \sim 1 \, {\rm pc},
\label{3}
\end{equation}
which corresponds to the observed jet radii~\cite{1}. One might expect such a
picture to be also preserved for an external medium with pressure 
$P\sim B^2/8\pi$; therefore, it seems of interest to consider the internal
structure of a one-dimensional jet immersed in an external uniform
magnetic field. However, a discussion of the more realistic case of a
medium with pressure is beyond the scope of this study. Nor do we
discuss the collimation itself but only consider the internal structure
of observed one-dimensional jets. This issue has become particularly
urgent because of the new possibilities offered by space radio
interferometry, which enables the internal structure of such jets to be
resolved. The effect of an external medium on the internal structure of
relativistic jets in the MHD model discussed here was previously studied
only by Appl and Camenzind~\cite{15, 16}. They considered only a special case
with a constant angular velocity of the plasma, in which the solution
with a zero total electric field flowing inside the jet could not be
constructed. As we show below, it is for the case of an angular velocity
decreasing toward the jet periphery (which, incidentally, is typical of
all models with a magnetic field passing through the accretion disk)
that the solution with finite magnetic flux $\Psi_0$ and zero total
current $I(\Psi_0) =0$ can be constructed.

On the other hand, many authors~\cite{22,23,24,25} obtained a universal solution
with a central core for a cylindrical jet:
\begin{equation}
B_z = \frac{B_0}{1+\varpi^2/\varpi_c^2},
\label{4}
\end{equation}
where, in the relativistic case,
\begin{equation}
\varpi_c = \frac{c\gamma}{\Omega}
\label{5}
\end{equation}
is the size of the central core, $\Omega$ is the angular velocity of the
compact object, and $\gamma$ is the characteristic Lorentz factor of the
outflowing plasma. As can be easily seen, such a solution results in a
rapid falloff of the poloidal field $B_z \propto \varpi^{-2}$ far from the rotation
axis  $\varpi\gg\varpi_c$. However, this solution is in conflict with the force-free
approximation, in which the poloidal magnetic field remains essentially
constant~\cite{14}. Indeed, when the energy density of the electromagnetic
field exceeds appreciably the plasma energy density (and it is this case
that was considered), it would be natural to assume that the internal
jet structure must be similar to the force-free one.

The examples given above show that a more detailed study with allowance
for all possible solutions is required even in the simplest case of a
one-dimensional cylindrical jet considered in terms of ideal
magnetohydrodynamics. Our study aims at a consistent investigation of
this issue.

Several features that we use when studying the structure of relativistic
jets typical of AGNs and radio pulsars should be immediately noted.
First of all, the jet radius $r_{\rm j}$ in all real cases proves to be
considerably larger than the light-cylinder radius $R_{\rm L} =c/\Omega$. This
implies that, when the internal structure of jets is investigated, the
corresponding equations must be written in complete relativistic form.
On the other hand, the gravitational forces can be disregarded in them
far from the compact object. Finally, for simplicity, we consider below
a cold plasma, which is justifiable because the thermal processes in the
magnetospheres of radio pulsars play no crucial role. As for the jets
from AGNs, this approximation is applicable here in those magnetospheric
regions in which the plasma density is low. In any case, this is true
for the field lines passing through the surface of a black hole.

The one-dimensional solutions describing collimated jets are obtained in
Sect. 2; analytic and numerical solutions for the basic physical
quantities characterizing the structure and physics of jets are given in
Sect. 3. The problem is solved in straightforward statement; i.e., all
jets characteristics are determined by a set of parameters in the
compact source and, most importantly, by the physical conditions in the
external medium. As a result, we have found the conditions under which
most of the energy in actual relativistic jets must be transferred by
the electromagnetic field, while a region with subsonic flow exists in
the central jet regions. We also show that the solution with a central
core~(\ref{4}) and~(\ref{5}) cannot be realized in an external magnetic field.
Finally, some astrophysical implications of the theory developed for
one-dimensional jets are discussed in Sect. 4.


\section{Basic equations}

Let us consider the structure of a one-dimensional jet where all
quantities depend only on radius $\varpi$; in what follows, the temperature of
the matter is assumed to be zero, and $c=1$. As in the general
axisymmetric case, it is convenient to describe the magnetic-field
structure in terms of magnetic-flux function $\Psi(\varpi)$, which is related
to the longitudinal magnetic field by
\be
B_z(\varpi)=\frac{1}{2\pi\varpi}\frac{d\Psi}{d\varpi}.
\label{6}
\ee

Accordingly, it is convenient to write the toroidal magnetic field, the
electric field, and the 4-velocity vector of the matter as
\be
B_\varphi(\varpi)=-\frac{2I}{\varpi},
\label{7}
\ee
\be
{\bf E}=-\frac{\Omega_{\rm F}}{2\pi}\frac{d\Psi}{d\varpi}{\bf e_\varpi},
\label{8}
\ee
\be
{\bf u}=\frac{\eta}{n}{\bf B}+\gamma\Omega_{\rm F}\varpi{\bf e_\varphi},
\label{9}
\ee
where $I(\varpi_0)$ is the total current within $\varpi < \varpi_0$. In the 
case of a cold plasma, at the cylindrical magnetic surfaces $\Psi={\rm const}$, four
"integrals of motion"~\cite{26} can be introduced, which should be considered
precisely as functions of magnetic flux $\Psi$ in the most general
statement.  These are primarily $\Omega_{\rm F}(\Psi)$ and $\eta(\Psi)$ in the
definitions~(\ref{8}) and~(\ref{9}), as well as the $z$-component of angular momentum
$L(\Psi) =I/2\pi+\mu\eta\varpi u_\phi$ and the energy flux 
$E(\Psi) =\Omega_{\rm F}I/2\pi+\gamma\mu\eta$. Here, $\mu$ is the
relativistic specific enthalpy, which is equal to the mass of particles
for a cold plasma. The specific form of the integrals of motion must be
determined from boundary conditions in the compact source and from
critical conditions at the singular surfaces.

As a result, the equilibrium equation for magnetic surfaces far from
gravitating bodies (Grad--Shafranov's equation) can be written as (see,
e.g.,~\cite{27})
\beq
\frac{1}{\varpi}\frac{d}{d\varpi}
\left(\frac{A}{\varpi}\frac{d\Psi}{d\varpi}\right)
+\Omega_{\rm F}(\nabla\Psi)^2\frac{d\Omega_{\rm F}}{d\Psi}
+\frac{64\pi^4}{\varpi^2}\frac{1}{2M^2}
\frac{d}{d\Psi}\left(\frac{G}{A}\right)
-\frac{32\pi^4}{M^2}\frac{d(\mu^2\eta^2)}{d\Psi}=0,
\label{10}
\eeq
where
$$
G=\varpi^2e^2+M^2L^2-M^2\varpi^2E^2,
$$
$$
A=1-\Omega_{\rm F}^2\varpi^2-M^2,
$$
$$
e(\Psi)=E(\Psi)-\Omega_{\rm F}(\Psi)L(\Psi).
$$
Here, $M^2 ={\bf u}_p^2/{\bf u}_a^2$ is the square of the Mach number with respect to the
Alfven velocity $u_a =B_z/(4\pi n\mu)^{1/2}$, and the derivative $d/d\Psi$ acts only on the
integrals of motion. The remaining jet parameters are given by the
well-known algebraic relations (see, e.g.,~\cite{27}):
\begin{eqnarray}
 \frac{I}{2\pi} & = & \frac{L-\Omega_{\rm F}\varpi^{2}E}
  {1-\Omega_{\rm F}^{2}\varpi^{2}-M^{2}},
\label{11} \\
\gamma & = & \frac{1}{\mu\eta}\,\frac{(E-\Omega_{\rm F}L)-M^{2}E}
{1-\Omega_{\rm F}^{2}\varpi^{2}-M^{2}},
\label{12} \\
 u_{\varphi} & = & \frac{1}{\varpi\mu\eta}\,\frac{(E-\Omega_{\rm F}L)
 \Omega_{\rm F}\varpi^{2}-LM^{2}}{1-\Omega_{\rm F}^{2}\varpi^{2}-M^{2}}.
\label{13}
\end{eqnarray}

Equation~(\ref{10}) contains four integrals of motion; this equation has no
singularity at the fast magnetosonic surface, because it depends only on
coordinate $\varpi$. As for the Alfven surface, $A=0$, the problem of the
boundary conditions generally requires a further study beyond the theory of ideal
magnetohydrodynamics. At the same time, for the fairly large currents 
$I\approx I_{\rm GJ}$ considered here, a solution continuous across this surface can 
always be constructed by a small change in the integrals of motion near the 
Alfven surface, $\Psi\approx\Psi_{\rm A}$. Consequently, equation~(\ref{10}) requires six boundary
conditions. These boundary conditions primarily include the external
uniform magnetic field
\be
B_z(r_{\rm j})=B_{\rm ext},
\label{14}
\ee
and the regularity condition at the magnetic axis $\varpi\to 0$
\begin{equation}
\Psi(\varpi)\rightarrow C\varpi^2.
\label{15}
\end{equation}

In addition, all four integrals $\Omega_{\rm F}$, $E$, $L$, and $\eta$ must be
specified. As for the remaining quantities characterizing the flow,
such as the jet radius $r_{\rm j}$ and the outflowing plasma
energy, they must be determined as a solution of the problem formulated
above. Similarly, the solution of the problem must also give an answer
to the question of whether the flow in the jet is supersonic.

Let us now consider the determination of the integrals of motion in more
detail. It would be natural to assume that, at the jet boundary where
there is no longitudinal motion of the matter, all four integrals of
motion become zero
\beq
\Omega_{\rm F}(\Psi_0) = 0, \quad E(\Psi_0) = 0, \quad 
L(\Psi_0) = 0, \quad \eta(\Psi_0) = 0.
\label{16}
\eeq
Here, $\Psi_0$ is the finite total magnetic flux concentrated in the
jet. This case corresponds to the absence of tangential discontinuities
at the jet boundary; according to~(\ref{11}), the total electric current
within the jet is automatically equal to zero.

We use the integrals of motion $\Omega_{\rm F}(\Psi)$, $L(\Psi)=I(\Psi)/2\pi$, 
and $E(\Psi)=\Omega_{\rm F}(\Psi)L(\Psi)$ derived by Beskin {\it et al.}~\cite{28} 
for the force-free magnetosphere of a
black hole, which satisfy the conditions~(\ref{16}) and, consequently, can be
directly used to study the jet structure. The only but very important change here is
the fact that, for a finite magnetization parameter $\sigma$~\cite{29},
\be
\sigma=\frac{\Omega_{\rm F}^2(0)}{8\pi^2}\frac{\Psi_0}{\mu\eta(0)},
\label{17}
\ee
which tends to infinity in the force-free approximation, the particle
contribution must be added to the energy integral $E(\Psi)$, because the
energy flux of the electromagnetic field near the rotation axis must
inevitably vanish. As a result, we have with accuracy up to $\sim\sigma^{-1}\ll 1$
\begin{equation}
\Omega_{\rm F}(\Psi)=
\frac{2\sqrt{1-\Psi/\Psi_0}}{1+\sqrt{1-\Psi/\Psi_0}}\Omega_{\rm F}(0),
\label{18}
\end{equation}
\begin{equation}
I(\Psi)=
\frac{1}{2\pi^2}\,\frac{\sqrt{1-\Psi/\Psi_0}}{1+\sqrt{1-\Psi/\Psi_0}}
\Omega_{\rm F}(0)\Psi,
\label{19}
\end{equation}
\begin{equation}
E(\Psi)=\gamma_{\rm in}\mu\eta+\Omega_{\rm F}(\Psi)L(\Psi).
\label{20}
\end{equation}
Below, we assume, for simplicity, that
\begin{equation}
\gamma_{\rm in} = {\rm const}.
\label{21}
\end{equation}

We emphasize that $\gamma_{\rm in}$ in expression~(\ref{20}) has the meaning of
the injection Lorentz factor in the region of the compact object and it is not equal
to the Lorentz factor of the jet particles.

Thus, we see from the formula for the energy flux $E(\Psi)$ that the
contribution by the electromagnetic field becomes dominant only at
$\Psi > \Psi_{\rm in}$, where
\begin{equation}
\Psi_{\rm in}=\frac{\gamma_{\rm in}}{\sigma}\Psi_0.
\label{22}
\end{equation}

At low values of $\Psi$, most of the energy is transferred by the
relativistic particles; as directly follows from relation~(\ref{20}), their
Lorentz factor is constant and equal to their initial value $\gamma_{\rm in}$. As for
the integral $\eta(\Psi)$, the particle-to-magnetic flux ratio, we chose it
in the form
\be
\eta(\Psi)=\eta_0(1-\Psi/\Psi_0),
\label{23}
\ee
which satisfies the condition~(\ref{16}).

We emphasize that the very possibility of using the integrals of motion
obtained by analyzing the inner magnetospheric regions, is not trivial.
Indeed, the flow outside the fast magnetosonic surface is completely
determined by four boundary conditions at the surface of a rotating
body. At the same time, a one-dimensional flow can be produced by the
interaction with the external medium, which gives rise (see, e.g.,~\cite{30})
to perturbations or shock waves propagating from "acute angles" and
other irregularities. Therefore, in regions where the conditions
for the validity of ideal magnetohydrodynamics are
violated, a significant redistribution of energy $E$ and angular momentum
$L$ is possible (e.g.~a part of them can be lost
via radiation). Nevertheless, we assume here, for simplicity,
that the integrals of motion $E(\Psi)$ and $L(\Psi)$, functions of flux
$\Psi$, remain exactly the same as those in the inner magnetospheric
regions.

In the one-dimensional case we consider, it is convenient to
reduce the second-order equation~(\ref{10}) to a set of two first-order
equations for $\Psi(\varpi)$ and $M^2(\varpi)$. Multiplying equation~(\ref{10}) by 
$2A(d\Psi/d\varpi)$, we obtain
\beq
\frac{d}{d\varpi}
\left[\frac{A^2}{\varpi^2}\left(\frac{d\Psi}{d\varpi}\right)^2\right]
+A\left(\frac{d\Psi}{d\varpi}\right)^2\frac{d'\Omega_{\rm F}^2}{d\varpi}
+\frac{64\pi^4 A}{\varpi^2 M^2}\frac{d'}{d\varpi}\left(\frac{G}{A}\right)
-\frac{64\pi^4 A}{M^2}\frac{d}{d\varpi}(\mu^2\eta^2)=0,
\label{24}
\eeq
with the derivative $d'/d\varpi$ acting only on the integrals of motion.
Finally, we use "Bernoulli's relativistic equation" $\gamma^2-{\bf u}^2=1$, which,
given the definitions of the integrals of motion $E(\Psi)$ and $L(\Psi)$, can be
written as
\beq
A^2\left(\frac{dy}{dx}\right)^2=
\frac{e^2}{\mu^2\eta^2}\frac{x^2(A-M^2)}{M^4}+
\frac{x^2E^2}{\mu^2\eta^2}-
\frac{\Omega_{\rm F}^2(0)L^2}{\mu^2\eta^2}-\frac{x^2A^2}{M^4},
\label{25}
\eeq
where we introduced the dimensionless variables
\begin{eqnarray}
x & = & \Omega_{\rm F}(0)\varpi=\Omega \varpi,
\label{26} \\
y & = & \sigma\Psi/\Psi_0.
\label{27}
\end{eqnarray}

As a result, substituting the right-hand part of~(\ref{25}) into the first term
of~(\ref{24}) and performing differentiation, we obtain the first first-order 
differential equation
\begin{eqnarray}
\left [\frac{e^2}{\mu^2\eta^2}+
\frac{\Omega_{\rm F}^2}{\Omega_{\rm F}^2(0)}x^2-1\right ]\frac{dM^2}{dx}=
\frac{M^6}{x^3A}\frac{\Omega_{\rm F}^2(0)L^2}{\mu^2\eta^2}-
\frac{xM^2}{A}\frac{\Omega_{\rm F}^2}{\Omega_{\rm F}^2(0)}
\left(\frac{e^2}{\mu^2\eta^2}-2A\right )
\nonumber \\
+\frac{M^2}{2}\frac{dy}{dx}\left[
\frac{1}{\mu^2\eta^2}\frac{de^2}{dy}+
\frac{x^2}{\Omega_{\rm F}^2(0)}\frac{d\Omega_{\rm F}^2}{dy}-
2\left(1-\frac{\Omega_{\rm F}^2}{\Omega_{\rm F}^2(0)}x^2\right)
\frac{1}{\eta}\frac{d\eta}{dy}\right].
\label{28}
\end{eqnarray}

The second first-order differential equation is Bernoulli's equation~(\ref{25}), which 
should now be considered as an equation for the derivative $dy/dx$. The set of
equations~(\ref{25}) and~(\ref{28}) allows a general solution to be constructed for
a one-dimensional jet immersed in an external magnetic field.

We emphasize one important advantage of the set of first-order equations~(\ref{25}) 
and~(\ref{28}) over the initial second-order equation~(\ref{10}). The point is
that the relativistic equation~(\ref{10}), which is basically the force
balance equation, contains the electromagnetic force
\be
{\bf F}_{\rm em} = \rho_{\rm e}{\bf E} + {\bf j} \times {\bf B},
\label{29}
\ee
in which the electric and magnetic contributions virtually cancel each
other far out from the rotation axis $\varpi\gg R_{\rm L}$. Using Bernoulli's
equation~(\ref{25}), we can derive~\cite{31}
\be
\frac{|\rho_{\rm e}{\bf E} + {\bf j} \times {\bf B}|}
{|{\bf j} \times {\bf B}|} \sim \frac{1}{\gamma^2}.
\label{30}
\ee

When analyzing~(\ref{10}), we therefore must retain all higher order terms 
$\sim\gamma^{-2}$, while the zero-order quantities 
$\rho_e{\bf E}$ and ${\bf j}\times{\bf B}$ in~(\ref{28}) are analytically 
removed using Bernoulli's equation, so all terms
of this equation are of the same order.

Finally, it is also important that the exact equation~(\ref{28}) has no
singularity near the rotation axis. In other words, its solution
contains no $\delta$-shaped current $I\propto\delta(\varpi)$ flowing along the
jet axis; several authors pointed out to the necessity of it~\cite{7,32}.


\section{Exact solutions and numerical results}

Let us consider the basic properties of the set of equations~(\ref{25}) and~(\ref{28}). 
As can be easily verified, in the relativistic case under
consideration, we may assume $\gamma= u_z$ with high accuracy. Far
from the rotation axis, $x\gg \gamma_{\rm in}$ ($\varpi\gg \gamma_{\rm in}R_{\rm L}$), 
equation~(\ref{25}) can be rewritten in the limit $M^2\ll x^2$ as 
\begin{equation}
\frac{d\Psi}{d\varpi}=
\frac{8\pi^2 E(\Psi)}{\varpi\Omega_{\rm F}^2(\Psi)}
\label{31}
\end{equation}
or, equivalently,
\begin{equation}
B_{z}(\varpi)=\frac{4\pi E(\Psi)}{\varpi^2\Omega_{\rm F}^2(\Psi)}.
\label{32}
\end{equation}

As we see, equation~(\ref{31}) does not contain $M^2$ at all and can therefore
be integrated independently. This must be the case, because equation~(\ref{31}) 
must coincide with the asymptotics of the force-free equation,
which can be derived from~(\ref{25}) by going to the limit $M^2 \to 0$. Assuming
now that $B_z(r_{\rm j})= B_{\rm ext}$ in~(\ref{32}), we obtain, in particular, for the jet
radius
\begin{equation}
r_{\rm j}^2= \lim_{\Psi\rightarrow\Psi_0}\frac{4\pi
E(\Psi)}{\Omega_{\rm F}^2(\Psi)B_{\rm ext}}.
\label{33}
\end{equation}

Consequently, the jet radius is determined by the limit of the 
$E(\Psi)/\Omega_{\rm F}^2(\Psi)$ ratio as $\Psi\to\Psi_0$. 
In particular, for $E(\Psi)$ and $\Omega_{\rm F}(\Psi)$ given {by~(\ref{18})--(\ref{20})},
we have
\begin{equation}
\lim_{\Psi\to\Psi_0}\frac{E(\Psi)}{\Omega_{\rm F}^2(\Psi)}
=\frac{1}{4\pi^2}\Psi_0,
\label{34}
\end{equation}
so the limit~(\ref{33}) does actually exist. As a result, we obtain
\begin{equation}
r_{\rm j}=\sqrt{\frac{\Psi_0}{\pi B_{\rm ext}}},
\label{35}
\end{equation}
which essentially coincides with estimate~(\ref{2}). This is no surprise, because
we show below that equations~(\ref{25}) and~(\ref{28}) for the integrals of
motion~(\ref{18}),~(\ref{20}), and~(\ref{23}) have a constant magnetic field as their solution
over a wide range of $\varpi$.

Let us now consider in more details the behavior of the solution in the
inner jet region, where $\Psi\ll \Psi_0$ and, hence, the integrals of motion can
be approximately written as
\begin{eqnarray}
L(\Psi) & = & \frac{\Omega_{\rm F}}{4\pi^2}\Psi,
\label{36} \\
\Omega_{\rm F}(\Psi) & = & \Omega={\rm const},
\label{37} \\
\eta(\Psi) & = & \eta_0={\rm const},
\label{38}
\end{eqnarray}
with $E(\Psi)=\gamma_{\rm in}\mu\eta_0+\Omega_{\rm F}L$ and 
$e(\Psi)= \gamma_{\rm in}\mu\eta_0={\rm const}$. As a result, 
$\Omega_{\rm F}L/\mu\eta_0=2y$, and we can rewrite equations~(\ref{25}) and~(\ref{28}) as
\begin{eqnarray}
{(1-x^2-M^2)}^2{\left (\frac{dy}{dx}\right )}^2=
 \frac{\gamma_{\rm in}^2 x^2}{M^4}(1-x^2-2M^2)+
x^2{(\gamma_{\rm in}+2y)}^2-4y^2-
 \frac{x^2}{M^4}{(1-x^2-M^2)}^2 ,
\label{39}
\end{eqnarray}
\begin{equation}
(\gamma_{\rm in}^2+x^2-1)\frac{dM^2}{dx}=2x M^2-
\frac{\gamma_{\rm in}^2 x M^2}{(1-x^2-M^2)}
+\frac{4y^2M^6}{x^3(1-x^2-M^2)}.
\label{40}
\end{equation}

Equations~(\ref{39}) and~(\ref{40}) describing the internal jet structure can be
solved analytically. It can be verified by direct substitution that we
have the following asymptotics for $x\ll\gamma_{\rm in}$:
\be
M^2(x) = M_0^2 = {\rm const},
\label{41}
\ee
\be
y(x) = \frac{\gamma_{\rm in}}{2M_0^2}x^2,
\label{42}
\ee
which correspond to a constant magnetic field
\be
B_{z}=B_{z}(0)=\frac{4\pi\gamma_{\rm in}\mu\eta_0}{M_0^2} =
\frac{\gamma_{\rm in}}{\sigma M_0^2}B(R_{\rm L}) = {\rm const},
\label{43}
\ee
where $B(R_{\rm L})=\Psi_0/R_{\rm L}^2$. Here, we assume that $\gamma_{\rm in}\gg 1$, 
which is the typical for jets from AGNs and radio pulsars.

The solution of~(\ref{39}) and~(\ref{40}) for $x\gg\gamma_{\rm in}$, i.e., at 
$\varpi\gg \gamma_{\rm in}R_{\rm L}$,
depends on the relationship between $\gamma_{\rm in}$ and 
$M_0=M(0)$. For example, at $M_0^2 > \gamma_{\rm in}^2$, when, according to~(\ref{43}), 
the axial magnetic field is fairly weak, the total magnetic flux 
within $\varpi < \gamma_{\rm in}R_{\rm L}$
\begin{equation}
\Psi(\gamma_{\rm in}R_{\rm L})\approx
\pi \gamma_{\rm in}^2R_{\rm L}^2B_z(0)
\label{44}
\end{equation}
can be written as
\begin{equation}
\Psi(\gamma_{\rm in}R_{\rm L})\approx
\frac{\gamma_{\rm in}^2}{M_0^2}\Psi_{\rm in},
\label{45}
\end{equation}
where the flux $\Psi_{\rm in}$ is given by~(\ref{22}). We see that, if the condition 
$M_0^2 > \gamma_{\rm in}^2$ is satisfied, then the total magnetic flux within 
$\varpi < \gamma_{\rm in}R_{\rm L}$ is lower
than $\Psi_{\rm in}$; so, outside this region, the particles also make the
main contribution to $E(\Psi)$ as before, while the contribution of the
electromagnetic field may be neglected. As a result, at 
$\varpi\gg \gamma_{\rm in}R_{\rm L}$,
the solution of~(\ref{39}) and~(\ref{40}) has a quadratic rise of $M^2$ and a
power-law falloff of the magnetic field~\cite{22, 23, 24}:
\be
M^2(x) = M_0^2\frac{x^2}{\gamma_{\rm in}^2}\gg x^2,
\label{46}
\ee
\begin{equation}
B_{z}(x) = B_z(0)\frac{\gamma_{\rm in}^2}{x^2}.
\label{47}
\end{equation}

Consequently, the magnetic flux increases very slowly (logarithmically)
with the distance from the rotation axis:
\begin{equation}
\Psi(x) \propto \ln(x/\gamma_{\rm in}).
\label{48}
\end{equation}

Such a behavior of the magnetic field, in turn, shows that the
transition flux $\Psi=\Psi_{\rm in}$ is reached exponentially far from the rotation
axis, in conflict with the estimate~(\ref{2}) corresponding to the assumption
of jet collimation.

We may thus conclude that an external constant magnetic field limits
the Mach number at the rotation axis above
\begin{equation}
M^2(0) < M_{\rm max}^2=\gamma_{\rm in}^2.
\label{49}
\end{equation}

Accordingly, as follows from~(\ref{43}), the magnetic field at the rotation
axis cannot be weaker than
\begin{equation}
B_{\rm min}= \frac{1}{\sigma\gamma_{\rm in}}B(R_{\rm L}).
\label{50}
\end{equation}

If, however, the Mach number at the rotation axis does not exceed $\gamma_{\rm in}$
(i.e.~if $M_0^2<\gamma_{\rm in}$),
then, as for the similar asymptotics $\varpi\ll\gamma_{\rm in}R_{\rm L}$, the solution of~(\ref{39})
and~(\ref{40}) for $\gamma_{\rm in}R_{\rm L}\ll\varpi\ll r_{\rm j}$ gives a constant magnetic 
field~(\ref{43}), which corresponds to the solution
\be
y(x) = \frac{\gamma_{\rm in}}{2M_0^2}x^2.
\label{51}
\ee

At the same time, in this case, we have only a linear increase in the
square of the Mach number
\be
M^2(x) = M_0^2\frac{x}{\gamma_{\rm in}}\ll x^2.
\label{52}
\ee

Then, according to~(\ref{27}) and~(\ref{51}), the jet radius can be written as
\begin{equation}
r_{\rm j}=\sqrt{\frac{\sigma M_0^2}{\gamma_{\rm in}}}R_{\rm L},
\label{53}
\end{equation}
which is equivalent to~(\ref{35}) [and in agreement with~(\ref{2})]. Moreover, as can be easily 
verified, the constant magnetic field $B_z=B(0)$ for the {invariants~(\ref{18})--(\ref{20})}
proves to be an exact solution of~(\ref{31}) in the entire jet
up to the jet boundary, $\varpi=r_{\rm j}$. Here, we may therefore assume $B(0)=B_{\rm ext}$.
Consequently, according to~(\ref{43}), we obtain
\be
M_0^2 = \frac{\gamma_{\rm in}}{\sigma}\frac{B(R_{\rm L})}{B_{\rm ext}}.
\label{54}
\ee

Using relation~(\ref{54}), we can also express all the remaining jet
parameters in terms of the external magnetic field.

Note that the absence of a declining solution $B_z \propto \varpi^{-2}$ [see~(\ref{4})] is
associated with the first term in the right-hand part of~(\ref{28})
proportional to $L^2$. This term, which changes appreciably the behavior
of the solution, appears to be missed previously. As it was already
emphasized above, this is not surprising because the corresponding term in the second-order
equation~(\ref{10}) is of high order and small. On the other
hand, far from the rotation axis $\varpi\gg\gamma_{\rm in}$, equation~(\ref{28}) can be
rewritten as
\be
\frac{d}{dx}
\left[\frac{\mu\eta\Omega_{\rm F}x^2}{M^2}\right]
+ \frac{\Omega_{\rm F}^{4}(0)M^2}{\Omega_{\rm F}\mu\eta x^3(x^2+M^2)}L^2 = 0,
\label{55}
\ee
in which both terms are of the same order. Neglecting
the term proportional to $L^2$, we arrive at the solution~(\ref{46}), $M^2\propto x^2$,
for $\Omega_{\rm F}={\rm const}$ and $\eta={\rm const}$. The conservation of function
\be
H = \frac{\mu\eta\Omega_{\rm F}x^2}{M^2}
\label{56}
\ee
was first found by Heyvaerts and Norman~\cite{7} for conical solutions, when
all quantities depend only on spherical coordinate $\theta$, but has
also been repeatedly discussed when analyzing cylindrical flows.
However, as we see, $H$ is generally not conserved in the cylindrical
geometry for relativistic jets. To be more precise, the second term in~(\ref{55}) 
turns out to be significant for all models with a nearly constant
density of the longitudinal electric current in the central jet region,
where the invariant $L(\Psi)$ linearly increases with magnetic
flux $\Psi$ if $\Psi\ll\Psi_0$.

Thus, we conclude that the solution with a central core~(\ref{4}) cannot be
realized in the presence of an external medium with a finite regular
magnetic field. This conclusion appears to be also valid in the presence
of a medium with finite pressure $P$. Indeed, since the magnetic flux~(\ref{48}) 
increases very slowly (logarithmically), the solution~(\ref{4}) yields an
exponentially large jet radius $r_{\rm j}\sim R_{\rm L}\exp{(\Psi_0/\Psi_{\rm in})}$. 
Accordingly, the magnetic
energy density must also be low at $\varpi\sim r_{\rm j}$. However, this configuration
cannot exist in the presence of an external medium with finite pressure
$P$, irrespective of whether it is produced by a magnetic field or by a
plasma. We may therefore conclude that the solutions with a central core
can be realized only for a special choice of the integral $L(\Psi)$, which
increases only slightly with the magnetic flux, and only in the absence of
an external medium. For the most natural (from our point of view) models
with a constant current density in the central jet regions, the solution
with a central core cannot be realized even in the absence of an
external medium.

In order to derive now the energy distribution in the jet and the
particle Lorentz factor, it is convenient to introduce the quantity
\be
g(x)=\frac{M^2}{x^2}.
\label{57}
\ee

Since at large distances $\varpi\gg \gamma_{\rm in}R_{\rm L}$, 
according to~(\ref{12}), we have
\be
\frac{\gamma\mu\eta}{E}=\frac{g}{g+1},
\label{58}
\ee
$g$ is simply the ratio of the energy flux transferred by particles
$W_{\rm part}$ to the energy flux of the electromagnetic field. As a result,
from relation~(\ref{52}) for $x\gg\gamma_{\rm in}$, we obtain
\be
\frac{W_{\rm part}}{W_{\rm tot}} \approx
\frac{M_0^2}{\gamma_{\rm in}}x^{-1}\ll 1,
\label{59}
\ee

Accordingly, from~(\ref{57}) and~(\ref{58}) for the particle Lorentz factor at 
$x\gg\gamma_{\rm in}$, we derive
\begin{equation}
\gamma(x)=x.
\label{60}
\end{equation}

Finally, expression~(\ref{13}) for the $\varphi$-component of the 4-velocity vector 
$u_\varphi$ yields the following toroidal velocity $v_\varphi= u_\varphi/\gamma$ 
at $x\gg \gamma_{\rm in}$:
\begin{equation}
v_{\varphi}(x) = \frac{1}{x}.
\label{61}
\end{equation}

We see that the particle energy approaches the universal asymptotic
limit~(\ref{60}) at $\varpi\gg \gamma_{\rm in}R_{\rm L}$. 
Naturally, such a simple asymptotics can also
be derived from simpler considerations. Indeed, using the ``frozen-in''
equation ${\bf E} + {\bf v}\times{\bf B}=0$, we obtain for the drift velocity
\begin{equation}
U_{\rm dr}^2=\frac{|{\bf E}|^2}{|{\bf B}|^2}
={\left (\frac{B_{\varphi}^2}{|{\bf E}|^2}+
\frac{B_z^2}{|{\bf E}|^2}\right )}^{-1}.
\label{62}
\end{equation}

In our case, however, according to~(\ref{6}) and~(\ref{7}), we have
\begin{equation}
B_{\varphi}^2 \approx |{\bf E}|^2,
\label{63}
\end{equation}
\begin{equation}
|{\bf E}|^2 \approx x^2B_z^2.
\label{64}
\end{equation}

As a result, {relations~(\ref{62})--(\ref{64})} immediately lead to the exact
asymptotics~(\ref{60}). It thus follows that, for example, for
electron--positron jets from AGNs, the jet particle energy is typically
\begin{equation}
{\bf \it E} \approx m_{\rm e}c^2\frac{r_j}{R_L}\sim (10^4 - 10^5)\,{\rm Mev}.
\label{65}
\end{equation}

On the other hand, according to~(\ref{57}), we reach a very important
conclusion that, far from the light cylinder, $\varpi\gg\gamma_{\rm in}R_{\rm L}$,
\begin{equation}
g \ll 1.
\label{66}
\end{equation}

Consequently, according to~(\ref{58}), the particle contribution to the
general energy flux balance proves to be minor. For example, at 
$B_{\rm ext}\sim B_{\rm min}$ for $r\sim r_{\rm j}$, we have
\begin{equation}
\frac{W_{\rm part}}{W_{\rm tot}}
\sim \sqrt{\frac{\gamma_{\rm in}}{\sigma}}\mbox{ .}
\label{67}
\end{equation}

While in the general case, we obtain
\begin{equation}
\frac{W_{\rm part}}{W_{\rm tot}} \sim
\frac{1}{\sigma}\left[\frac{B(R_{\rm L})}{B_{\rm ext}}\right]^{1/2}.
\label{68}
\end{equation}

Thus, we reach a fairly nontrivial conclusion that the fraction of
energy transferred by particles in a one-dimensional jet must be
determined by the parameters of the external medium.

\begin{figure}[p]
\vspace{20.0cm}
\includegraphics{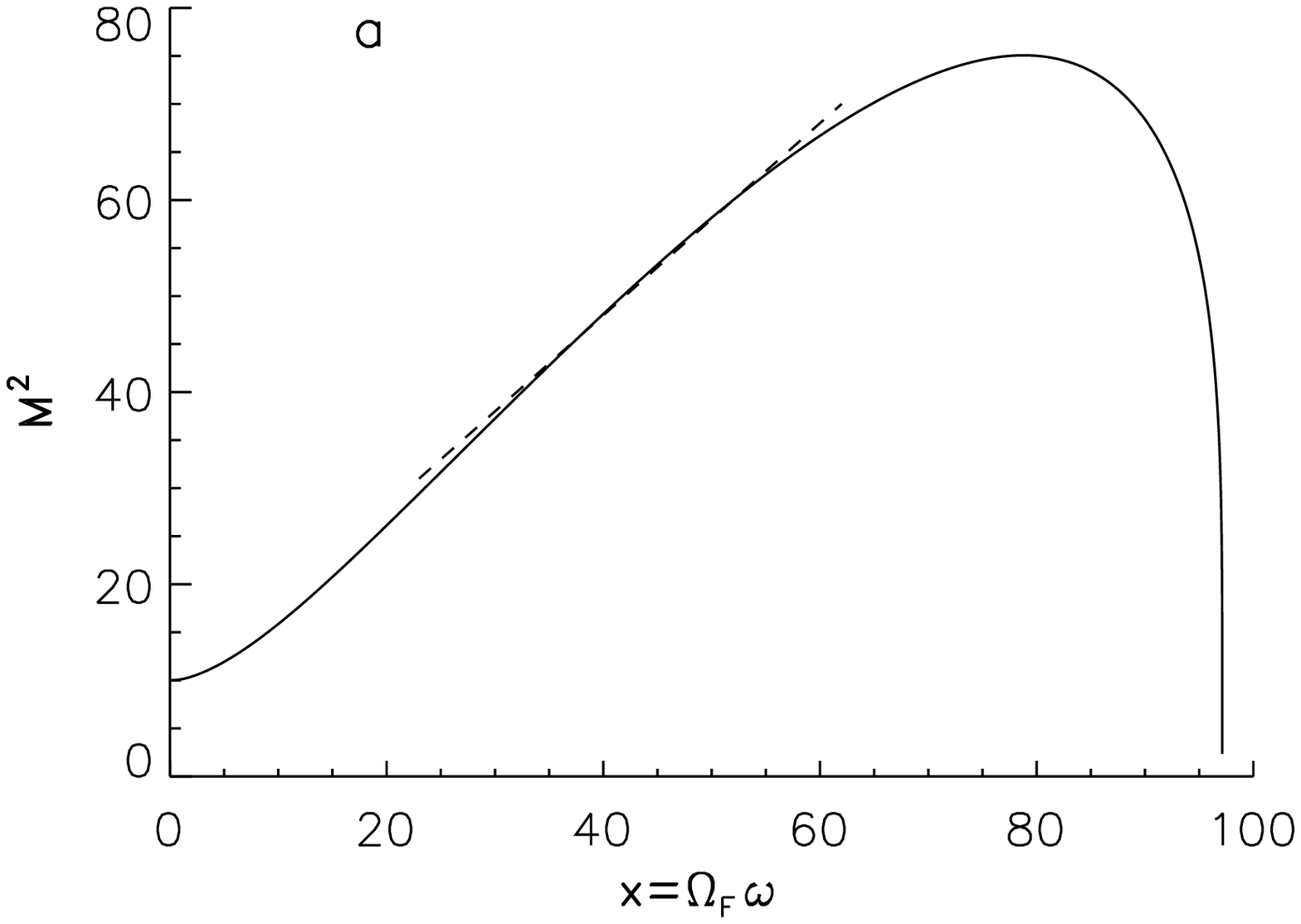}
\includegraphics{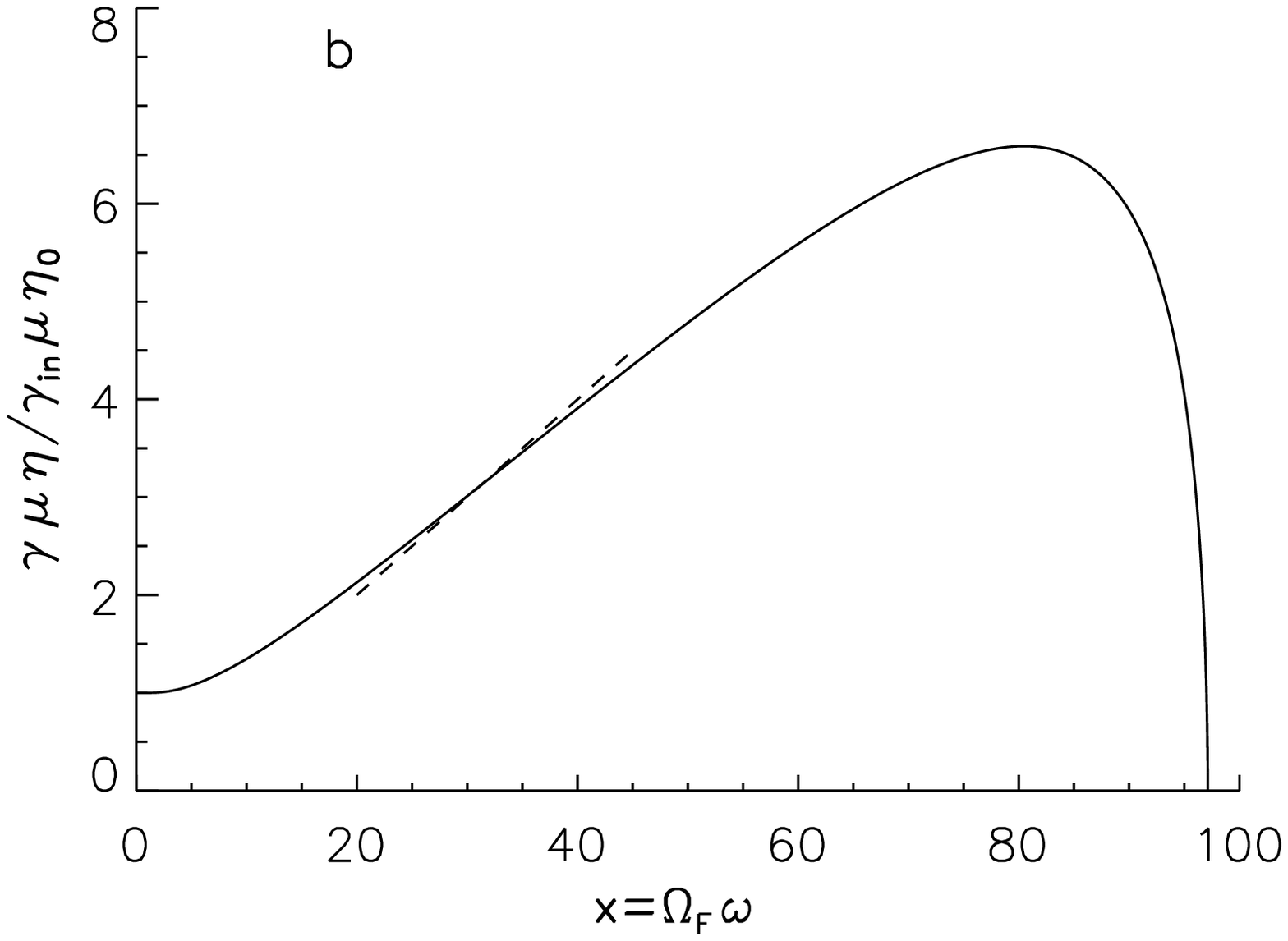}
\includegraphics{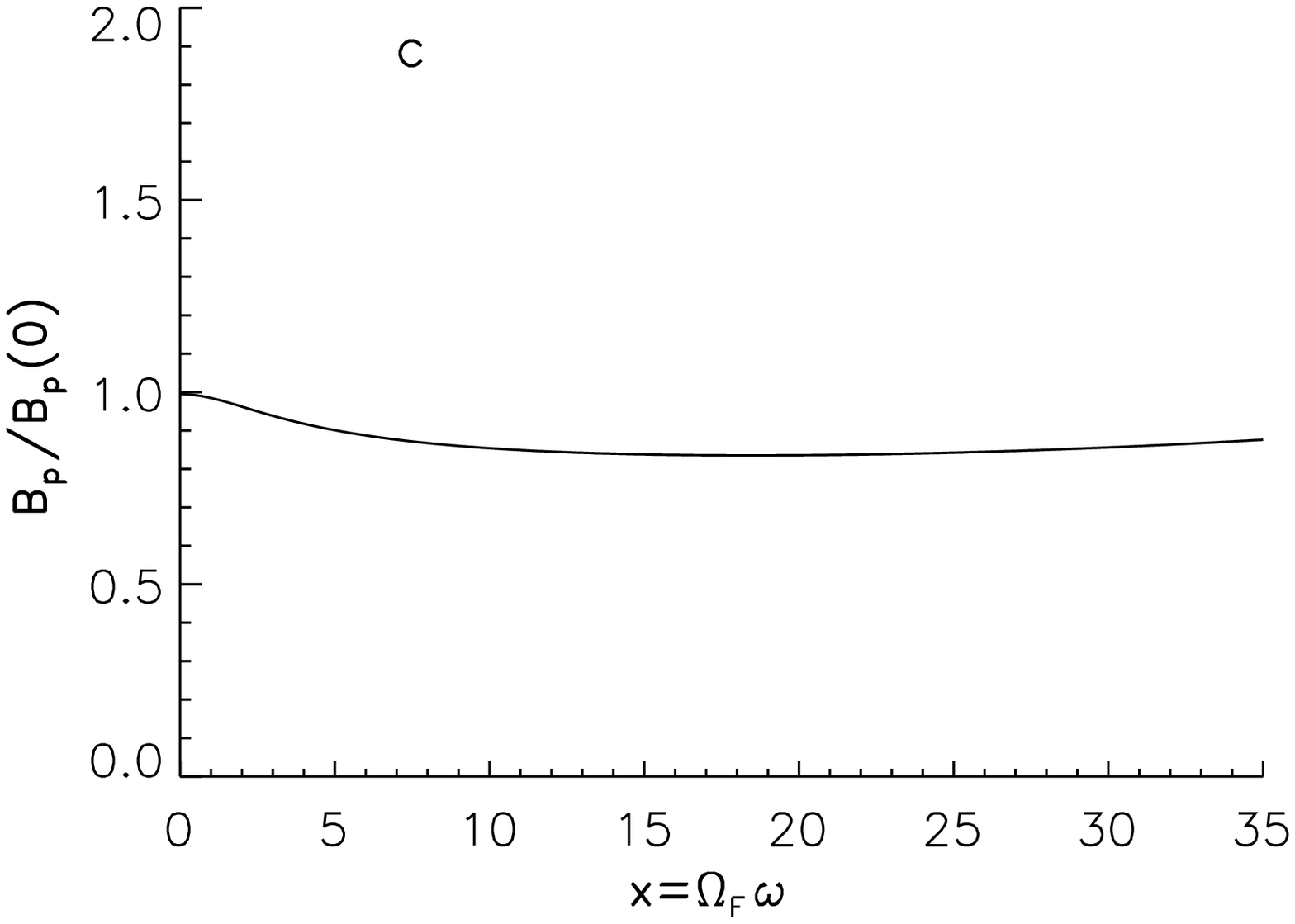}
\caption{The dependences of Mach number $M^2$ (a), energy flux 
$\gamma\mu\eta$ concentrated in particles (b), and poloidal magnetic field $B_z$ (c)
obtained by numerical integration of the equilibrium equations~(\ref{25}) 
and~(\ref{28}) for $M_0=16$, $\gamma_{\rm in}=8$, and $\sigma=1000$. The dashed lines
indicate the behavior of these quantities that follows from the analytic
asymptotics.}
\label{FIG_1}
\end{figure}

Let us now discuss the results of exact calculations obtained by
numerical integration of equations~(\ref{25}) and~(\ref{28}) with the integrals of
{motion~(\ref{18})--(\ref{20})} and~(\ref{23}). In figures~\ref{FIG_1}(a) and~\ref{FIG_1}(b), 
the Mach number and the energy flux concentrated in particles $\gamma\mu\eta$ are plotted against 
$x=\Omega_{\rm F}\varpi$ for $M_0^2=16$, $\gamma_{\rm in}=8$, and $\sigma=1000$.
The dashed lines indicate the
behavior of these quantities that follows from the analytical asymptotics~(\ref{52}) 
and~(\ref{60}). As we see, at sufficiently small $x$ when the integrals of
{motion~(\ref{18})--(\ref{20})} and~(\ref{23}) are similar {to~(\ref{36})--(\ref{38})}, 
the analytical asymptotics match the exact numerical results. On the other hand, as expected,
$\gamma\mu\eta$ and $M^2$ are zero at $\Psi=\Psi_0$, i.e., at the jet edge. 
Figure~\ref{FIG_1}(c) shows the dependence of the poloidal field, $x^{-1}(dy/dx)$, for the inner 
parts of the jet, $\Psi<\Psi_0$. We see from this figure that the magnetic field is
nearly constant at $x > \gamma_{\rm in}$, in agreement with the analytic 
estimates~(\ref{43}) and~(\ref{51}). Of course, in general, the structure of the poloidal magnetic 
field is determined by a specific choice of the integrals $E(\Psi)$ and
$L(\Psi)$.

In conclusion, it is of interest to compare the energy of jet particle with
the limiting energy acquired by the particles as they outflow from the
magnetosphere of a compact object with a monopole magnetic field.
According to calculations by Beskin et al.~\cite{33}, the particle Lorentz factor outside 
the fast magnetosonic surface $r > \sigma^{1/3}R_{\rm L}$ in the absence of 
an external medium can be written as
\begin{eqnarray}
\gamma(y) & = & y^{1/3}, \qquad y > \gamma_{\rm in}^3,
\label{69} \\
\gamma(y) & = & \gamma_{\rm in}, \qquad \quad y < \gamma_{\rm in}^3,
\label{70}
\end{eqnarray}
where $y$ is given by~(\ref{27}). On the other hand, relations~(\ref{51}) and~(\ref{60})
for the jet yield
\begin{eqnarray}
\gamma(y) & = & \left(\frac{M_0^2}{\gamma_{\rm in}}\right)^{1/2}y^{1/2},
\qquad y > \frac{\gamma_{\rm in}^3}{M_0^2},
\label{71} \\
\gamma(y) & = & \gamma_{\rm in}, \qquad \qquad \qquad
y < \frac{\gamma_{\rm in}^3}{M_0^2}.
\label{72}
\end{eqnarray}

\begin{figure}[p]
\vspace{14cm}
\includegraphics{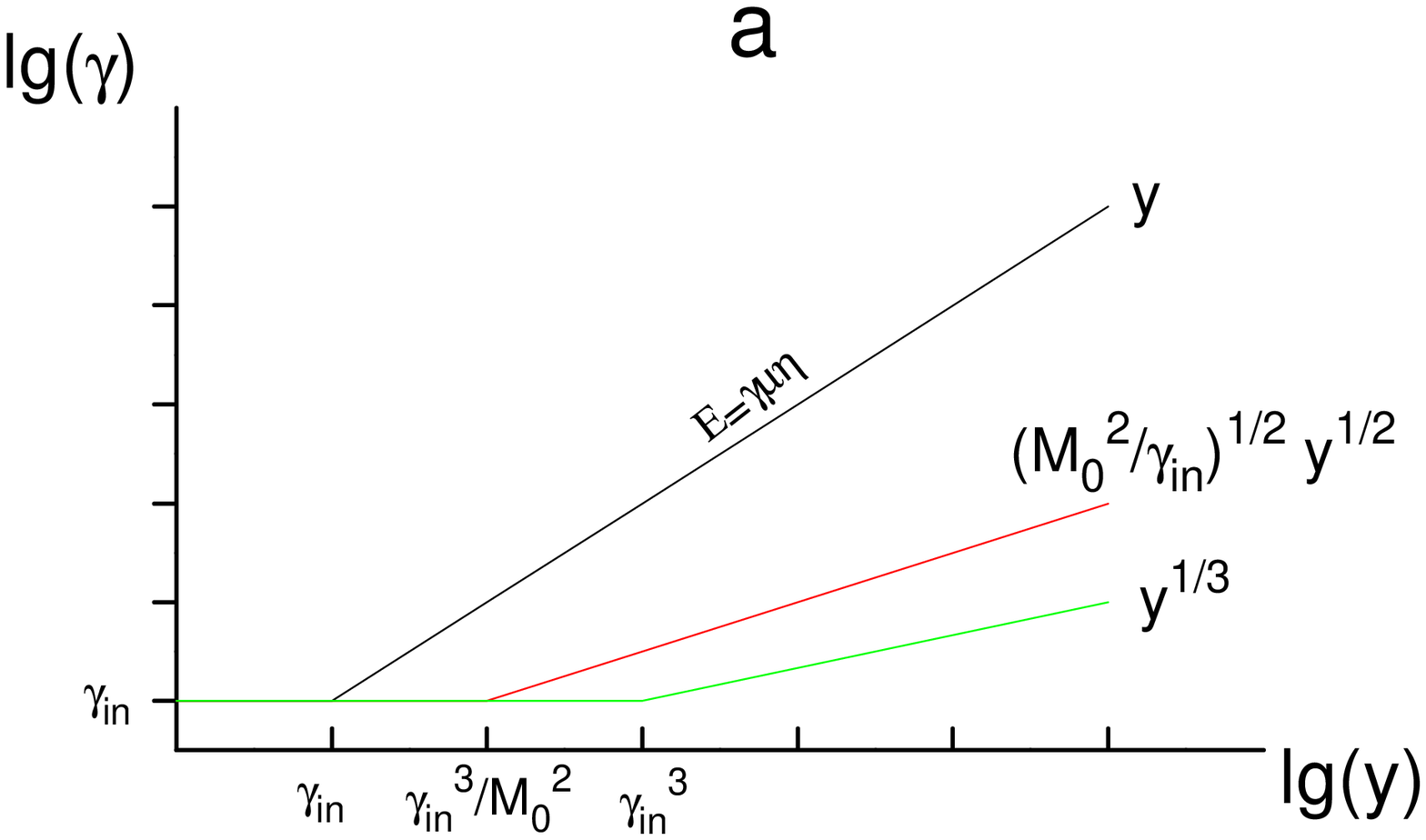}
\includegraphics{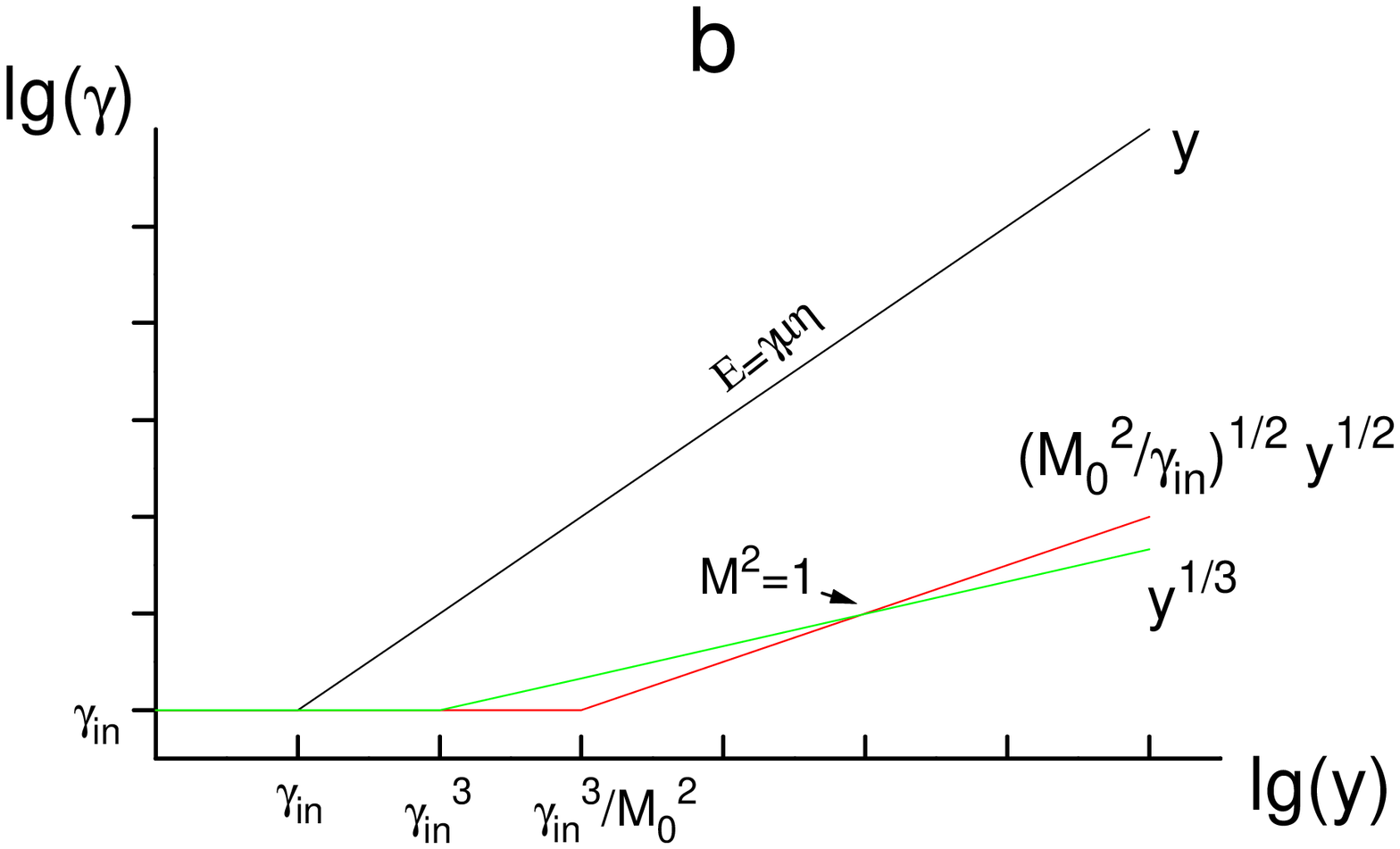}
\caption{Logarithmic dependence of the particle Lorentz factor on the
magnetic flux for a monopole field $\gamma\sim y^{1/3}$, for a
one-dimensional jet $\gamma\sim (M_0^2/\gamma_{\rm in})^{1/2}y^{1/2}$, 
and for complete transformation
of the electromagnetic energy into the particle energy $\gamma\sim y$ at:
(a) $M^2 > 1$ ($B_{\rm ext} < B_{\rm cr}$); and (b) $M^2 < 1$ ($B_{\rm ext} > B_{\rm cr}$). 
The intersection of the straight lines ($M^2\approx 1$) determines the position of
the fast MHD point.}
\label{FIG_2}
\end{figure}

As shown in figure~\ref{FIG_2}(a), for $M^2 > 1$, i.e., for $B_{\rm ext} < B_{\rm cr}$, where
\begin{equation}
B_{\rm cr} = \frac{\gamma_{\rm in}}{\sigma}B(R_{\rm L}),
\label{73}
\end{equation}
the Lorentz factor of the jet particles~(\ref{71}) is always larger than the
Lorentz factor acquired by the particles as they outflow from a
magnetosphere with a monopole magnetic field, but, of course, is always
smaller than the critical Lorentz factor
\begin{equation}
\gamma(y)=y,
\label{74}
\end{equation}
which corresponds to the complete transformation of the electromagnetic
energy into the particle energy. This implies that, at $B_{\rm ext} < B_{\rm cr}$, the
particles must be additionally accelerated during the collimation
coupled with the interaction of the outflowing plasma with the external
medium. If, alternatively, $B_{\rm ext} > B_{\rm cr}$, then, in the inner jet regions,
at
\begin{equation}
\varpi < \frac{\gamma_{\rm in}}{M_0^2}R_{\rm L},
\label{75}
\end{equation}
the particle energy on a given field line turns out to be even lower
than that for a monopole magnetic field, as shown in figure~\ref{FIG_2}(b).

The latter result can be easily explained. Indeed, for the integrals of
{motion~(\ref{36})--(\ref{38})} we consider, the factor $D$, whose zero value
determines the location of the fast MHD surface (see~\cite{27} for more
detail), can be rewritten in the case of a cold plasma as
\begin{eqnarray}
M^{2}D\equiv A+\frac{B_\varphi^2}{B_{\rm p}^2} =
A+\frac{4x^2y^2M^4}{4y^2M^4-2x^2M^2-x^4}.
\label{76}
\end{eqnarray}

It is easy to show that expression~(\ref{76}) for $y$ and $M^2$ given by~(\ref{51}) 
and~(\ref{52}) is negative at
\begin{equation}
M^2 \sim 1,
\label{77}
\end{equation}
i.e., at $x$ corresponding to~(\ref{75}). Consequently, we may reach
another important conclusion that, for sufficiently strong external
magnetic fields $B_{\rm ext} > B_{\rm cr}$~(\ref{73}) when $M^2 < 1$, a region with a 
subsonic flow inevitably emerges in the inner jet regions $\varpi < r_s$, where
\begin{equation}
r_{\rm s} \approx
\sigma\left[\frac{B_{\rm ext}}{B(R_{\rm L})}\right]^{3/2}R_{\rm L}.
\label{78}
\end{equation}
At the same time, a region with the subsonic flow can be produced far from
the compact object either by a shock wave or by a strong distortion of
the magnetic field within the fast magnetosonic surface located in the
vicinity of the compact object. In both cases, the magnetic-field
perturbation causes the particle energy to decrease.


\section{Discussion}

Thus, we conclude that the exact equilibrium equations~(\ref{25}) and~(\ref{28})
do actually allow a construction of a self-consistent model for a jet
immersed in an external uniform magnetic field. The advantage of these equations over
equation~(\ref{10}) results from the fact that all terms in~(\ref{28}) are of the same
order. In this case, the uniformity of the poloidal
magnetic field within the jet~(\ref{43}) results from the choice of 
{integrals~(\ref{36})--(\ref{38})}. In general, the poloidal magnetic field depends 
on the specific form of the integrals. A full analysis of the possible
solutions is beyond the scope of this paper.

We have shown that the fraction of energy transferred by particles
$W_{\rm part}/W_{\rm tot}$ must be largely determined by the parameters of the
external medium. In case $\sigma >> \sigma_{\rm cr}$, where
\begin{equation}
\sigma_{\rm cr} = \left[\frac{B(R_{\rm L})}{B_{\rm ext}}\right]^{1/2},
\label{79}
\end{equation}
the energy transferred by particles is only a small fraction of the
energy flux $W_{\rm em}$ transferred by the electromagnetic field. Consequently,
the jet is strongly magnetized ($W_{\rm part}\ll W_{\rm tot}$) only at sufficiently
large $\sigma$. If, however, the magnetization parameter does not exceed
$\sigma_{\rm cr}$, then, in this case, an appreciable part of the energy in the jet is
transferred by particles. This, in turn, implies that a considerable
part of the energy must be transferred from the electromagnetic field to
the plasma particles during jet collimation. It is interesting that $\sigma_{\rm cr}$
turns out to be approximately the same both for AGNs and for fast radio
pulsars:
\be
\sigma_{\rm cr} \approx 10^5 - 10^6.
\label{80}
\ee

We have shown that the central part of the jet must be subsonic for
sufficiently strong external magnetic fields. Thus, the theory gives
direct predictions whose validity can be verified by observations. It
should also be noted that the results obtained above are applicable both
to electron--positron and to electron--proton jets. However,
a direct evidence that the jets in AGNs are actually
electron--positron ones has recently appeared~\cite{34, 35}.

In our view, an important result is that, if the
external regular magnetic field is taken into account, the MHD equations
allow a self-consistent model to be constructed for a jet with a zero
total longitudinal electric current, $I(\Psi_0)=0$. In this case, a uniform
magnetic field that matches the external magnetic field can also be a
solution for the inner jet regions. As was already emphasized above, the
radii of the jets from AGNs can thus be also explained in a natural way.
In addition, since only a small fraction of the electromagnetic-field
energy is transformed into the particle energy, the energy transfer from
the compact object in the region of energy release can be explained as
well. At the same time, extending the MHD solution to the jet region
requires very high particle energies ($\sim 10^4\,{\rm MeV}$), which have not
been recorded yet. However, a consistent discussion of the outflowing-plasma
energy requires a proper allowance for the particle interaction with the
surrounding medium (for example, with background radiation), which may
cause a significant change in particle energy.

As for the quantitative predictions about the real physical parameters
of jets, they, as we showed above, essentially depend only on the
following three quantities: the magnetization parameter $\sigma$~(\ref{17}),
the Lorentz factor $\gamma_{\rm in}$ in the generation region, and the external magnetic field
$B_{\rm ext}$. In this case, the main uncertainty for electron--positron jets
from AGNs ($B_{\rm in}\sim 10^4\,{\rm G}$, $R_{\rm in}\sim 10^{14}\,{\rm cm}$) is the value of 
the magnetization parameter. Indeed, this quantity depends on the efficiency
of pair production in the magnetosphere of a black hole, which, in turn,
is determined by the density of hard gamma-ray photons. As a result, if
the density of hard gamma-ray photons with energies $E_\gamma > 1\,{\rm MeV}$ near the
black hole is high enough, then the particles will be produced by direct
collisions of photons $\gamma+\gamma \to e^+ + e^-$~\cite{36}. This causes an abrupt
increase in the multiplicity parameter {$\lambda=n/n_{\rm GJ}\sim 10^{10}$--$10^{12}$}, where
$n_{\rm GL}\approx \Omega B/2\pi c$ is the characteristic particle density required to shield the
longitudinal electric field. Using the well-known estimate (see, e.g.,~\cite{31})
\begin{equation}
\sigma \sim \frac{\Omega e B_{\rm in} R_{\rm in}^2}{\lambda m_{e}c^3},
\label{81}
\end{equation}
we obtain
\begin{equation}
\sigma \sim 10 - 10^3. \qquad \gamma_{\rm in} \sim 3 - 10.
\label{82}
\end{equation}

On the other hand, for low densities of gamma-ray photons when an
electron--positron plasma can be produced only in regions with a nonzero
longitudinal electric field, which are equivalent to the outer gaps in the
magnetospheres of radio pulsars~\cite{37, 38}, the multiplicity of the particle
production is fairly small: {$\lambda\sim 10$--$100$}. In this case, we obtain
\begin{equation}
\sigma \sim 10^{11} - 10^{13}, \qquad \gamma_{\rm in} \sim 10.
\label{83}
\end{equation}

Finally, for fast Crab- or Vela-type radio pulsars ($B_{\rm in}\sim 10^{13}\,{\rm G}$,
polar-cap radius $R_{\rm in}\sim 10^5\,{\rm cm}$, and $\lambda= n/n_{\rm GJ}\sim 10^4$) 
in which jets are observed, we have~\cite{31}
\begin{equation}
\sigma \sim 10^{6} - 10^{7}, \qquad \gamma_{\rm in} \sim 10^{2} - 10^{3}.
\label{84}
\end{equation}

Relations~(\ref{82}) and~(\ref{83}) show that the properties of the jets from AGNs
considerably depend on the magnetization parameter $\sigma$. For example,
according to~(\ref{79}), the jet particles for sources with large 
$\sigma\sim 10^{12}$ transfer only a small fraction of the energy compared to the
electromagnetic flux, so the flow within the jet differs only slightly
from the force-free flow. In addition, in this case the external magnetic field 
$B_{\rm ext}\sim 10^{-6}\,{\rm G}$ exceeds the critical magnetic field 
$B_{\rm cr}\sim 10^{-7}\,{\rm G}$.
According to~(\ref{43}) and~(\ref{73}), this implies that a subsonic region must
exist in the inner regions of such jets. On the other hand, a
substantial part of the energy in sources with magnetization parameter
$\sigma\sim 100$ during jet collimation must be transferred by plasma
particles, and no subsonic region is formed near the rotation axis. As
for the fast radio pulsars, the condition $B_{\rm ext}\ll B_{\rm cr}$ is satisfied 
for them, so a subsonic region
in the central parts of pulsar jets is not achieved either. On the other hand,
the estimate~(\ref{79}) shows that an appreciable part of the jet total energy
must be coupled with particles.

We emphasize that, since the jet radius~(\ref{53}) for AGNs always exceeds the
light-cylinder radius by several orders of magnitude,
\begin{equation}
r_{\rm j} \sim (10^4 - 10^5)R_{\rm L},
\label{85}
\end{equation}
the toroidal magnetic field $B_\phi$ within the jet must exceed the
poloidal magnetic field $B_p$ in the same proportion,
\begin{equation}
B_{\varphi} \approx \frac{r_{\rm j}}{R_{\rm L}}B_{\rm p}
\sim (10^4 - 10^5) B_{\rm ext}.
\label{86}
\end{equation}

Consequently, detection of such a strong toroidal component would be a
crucial argument for the picture discussed here. Unfortunately,
determination of the actual physical conditions in jets currently
involves considerable difficulties. Nevertheless, not only data on the
direct detection of such a structure~\cite{39} but also evidence for the
existence of magnetic fields $B\sim 0.1\,{\rm G}$, closely matching the 
estimate~(\ref{86})~\cite{34}, have recently appeared.


\section*{Acknowledgments}

We wish to thank A.V. Gurevich for interest in this study, a useful
discussion, and support. We also wish to thank S.V. Bogovalov, L.I.
Gurvits, and S.A. Lamzin for fruitful discussions. This study was
supported in part by the INTAS grant no. 96--154 and the Russian
Foundation for Basic Research (project number 99-02-17184). L. Malyshkin is
also grateful to the International Science Foundation.





\end{document}